\documentclass[twocolumn,11pt]{article}
\usepackage{txfonts}
\usepackage{lscape}
\usepackage{epsfig}
\usepackage{amssymb}
\usepackage{txfonts}
\usepackage{graphicx}
\usepackage{natbib} 

\newcommand\ion[2]{#1{\scshape #2}}
\newcommand\sun{\odot}
\newcommand\degr{$^\circ$}

\begin{document} 

   \title{Keplerian rotation of our Galaxy?} 
   \author{ { P. Gnaci\'nski$^{1}$ and T. M\l ynik$^{2}$ }  \\ 
   $^{1}$Institute of Theoretical Physics and Astrophysics, \\
          Faculty of Mathematics, Physics and Informatics, \\
              University of Gda\'nsk, \\
              80-308 Gda\'nsk, Poland \\
              email: {\it fizpg@univ.gda.pl} \\ 
   $^{2}$Student at the University of Gda\'nsk.
   }



   \maketitle  
   
   \begin{abstract}  
      It is common to attribute a flat rotation curve to our Galaxy.
      However \cite{Gazinur} in a recent paper have obtained a Keplerian rotation curve for interstellar clouds
      in outer parts of the Galaxy. They have calculated the distances from equivalent widths of
      interstellar \ion{Ca}{II} lines. The radial velocity was also measured on the interstellar \ion{Ca}{II}
      absorption line.
      
      We verify the result by \cite{Gazinur} basing on observations of old open clusters.
      We propose, that the observations of flat and Keplerian rotation curves may be caused by the assumption of circular orbits.
      The application of formulas derived with the assumption of circular orbits to elliptical ones may
      mimics the flat rotation curve.
      The interstellar clouds with cross-sections larger than stars may have almost circular orbits,
      and the derived rotation curve will be Keplerian.

   \end{abstract}  
   
   {\bf  Key words: }{\it  
     Galaxy kinematics and dynamics; dark matter
   }  
   


\section{Introduction}

   Our Galaxy is usually thought to have a flat rotation curve. 
The flat rotation curves of galaxies are usually explained assuming existence of dark matter.
The MOND (MOdified Newtonian Dynamics) models are less popular. 
However, some galaxies have Keplerian rotation curves which falls as $\sim 1/\sqrt{r}$
in the outer parts of the galaxies. In a sample of 45 galaxies analyzed by \cite{Honma1997}
11 have Keplerian rotation curve.

  A compilation of rotation velocities observed in our Galaxy was made by \cite{Sofue2009}.
They have transformed the rotational velocities from various sources to
common parameters $R_{\sun}=8$ kpc and $v_{\sun}=200$ km/s.
 In this paper we adopted the recently obtained solar velocity $v_{\sun}=240$ km/s \citep{Honma2015,Honma2012,Sofue2016}.
 The rotation curves and rotational velocities of individual objects were recalculated using $v_{\sun}=240$ km/s.

Rotation velocities from \cite{Sofue2009}, derived with tangent
point method or from radial velocity, are shown on Fig. \ref{rotVel}(a).

  The absence of dark matter in solar neighborhood was postulated by \cite{Bidin}.
Their result is based on stellar kinematics in direction perpendicular to the galactic plane.
However, their calculation leads to a flat rotation curve.

  The recent paper by \cite{Gazinur} shows a Keplerian rotation curve of our Galaxy.
They have based on distances and radial velocities derived from interstellar \ion{Ca}{II} absorption lines.
The aim of this paper is to reconcile the flat rotation curve from \cite{Sofue2009} and the Keplerian
rotation derived by \cite{Gazinur}.

\section{Old open clusters}

\begin{figure*}
    \centering
    \includegraphics[width=\textwidth]{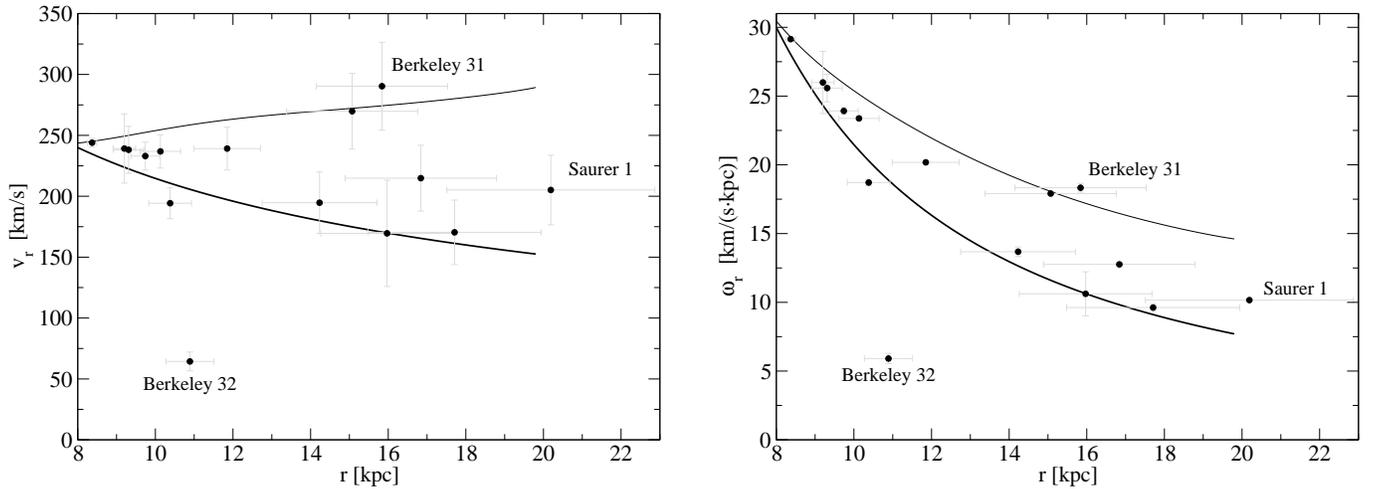}
    \caption{ Rotation velocity of old open clusters (left panel) versus Galactocentric distance. The right panel show the same data
      as angular rotation velocity. The lines show the model of flat rotation curve by \protect\cite{Sofue2009} and the Keplerian rotation curve (both for v$_\sun$=240~km/s). }
    \label{fig_OC}
\end{figure*}

   In order to verify the result by \cite{Gazinur} we
   have analyzed the rotation velocity of old open clusters (age greater than $10^9$ years)
   located in the outer part of the Galaxy $l\in(90^\circ, 270^\circ)$.
   All analyzed open clusters are located close to the galactic plane $|b|<20^\circ$.

   The rotational velocity was calculated from observed radial velocity. The heliocentric radial velocity
   $v_h$ was first transformed to the local standard of rest (LSR) 
   \begin{equation} 
     v_{LSR}=v_h + U_\sun \cos b \cos l + V_\sun \cos b \sin l + W_\sun \sin b               \label{v_LSR}
   \end{equation}
   using the Sun velocity $(U_\sun,V_\sun,W_\sun)=(11.1,12.24,7.25)$ km/s from \cite{Schonrich}.
   The rotational velocity was calculated using formula derived for circular orbits \citep[eg.][]{Bhattacharjee}
   \begin{equation} 
     v(r)=\frac{r}{R_{\sun}} \left( \frac{v_\mathrm{LSR}}{\sin l \cos b} + v_{\sun} \right).   \label{vr_zb}
   \end{equation}
   In this formula $r$ is the projection of galactocentric distance on the galactic plane
   \begin{equation} 
     r=\sqrt{R_\sun^2 + d^2 \cos^2 b -2R_\sun d \cos b \cos l}.             \label{r}
   \end{equation}
   The clusters with galactic longitude $l=180^\circ\pm20^\circ$ were excluded from our sample, 
   because the $\sin l$ in the denominator of formula \ref{vr_zb} leads to unphysical (i.e. negative) rotation velocities.
   
   At least some old open clusters have nearly circular orbits. The five old open clusters analyzed
   by \cite{Carraro1994} have eccentricities less than 0.14, with two clusters having eccentricities as low as $e=0.03$.
   We have collected open clusters data from the literature (see table \ref{tab_OC}), and determined the
   rotation velocity using formulas \ref{v_LSR} and \ref{vr_zb}. 
   The open clusters linear velocity, as well as the angular velocity is presented on figure \ref{fig_OC}.
   The advantage of the angular velocity is, that its error does not depend from the distance to cluster,
   which is known with little accuracy.
   
    The distances to open clusters analyzed by \cite{Carraro2007} were determined by fitting a
   isochrone to the CMD (colour--magnitude diagram). The largest error of distance in their sample
   of five open clusters is 21\%. The distance to Saurer~1 was also determined by fitting a
   isochrone to the CMD. For the open cluster Berkeley~31 we were unable to track down the method
   used to determine distances. The distance of 8.3~kpc to Berkeley~31 was cited by \cite{Carraro2007}, but
   other distances can be found in the literature: the distance 3.68~kpc was cited by \cite{Janes1994}, and 5.2$\pm$0.5~kpc
   was determined by fitting isochrones to the CMD \citep{Guetter1993}.
   Distances to other clusters were determined using the
   synthetic CMD method \citep{Tosi91}, but the errors of distances were not given. The authors state,
   that the synthetic CMD method is more accurate than the isochrone fitting to CMD. Therefore we
   have adopted the relative error of distances equal to 21\% for all analysed open clusters.
   
    The distances to open clusters are known with better accuracy, than the distances to \ion{H}{II}
   regions, which were used by \cite{Sofue2009} to construct his rotation curve.
   The distances to \ion{H}{II} regions were determined using optical spectrophotometric methods 
\citep{Fich1989}. The maximal relative error of their distances is 40\%, and the average error is 25\%. 

   Figure \ref{fig_OC} presents rotational velocity of old open clusters. The same data is presented as angular velocity, because
   angular velocity error does not depend on distance error. Therefore we have checked the agreement between open clusters velocity
   and flat/Keplerian rotation curves with the angular velocity data. Because errors of the angular velocity are negligible as compared to
   distance errors we have analyzed the data as a $r(\omega)$ function. We have computed 
   \begin{equation}
     \chi^2=\sum_i\left( \frac{r_i -r(\omega_i)}{\sigma_i}\right)^2 .
   \end{equation}
   For Keplerian rotation curve we got $\chi^2=30.4$ , while for the flat rotation we have $\chi^2=377.9$. 
   The angular velocity of analyzed open clusters agrees with the Keplerian rotation curve at the significance level $\alpha=0.005$.
   The open cluster Berkeley 32 was excluded from this analysis.

\begin{landscape}
\begin{table}
\centering
\begin{tabular}{lrrrrrrrrrrr}   
\hline         
Cluster	&	radial velocity	&	stars 	&	ref. 	&	 \multicolumn{1}{c}{l}	&	 \multicolumn{1}{c}{b}	&	\multicolumn{1}{c}{dist}	&	\multicolumn{1}{c}{Age}	&	ref. 	&	\multicolumn{1}{c}{r} &	\multicolumn{1}{c}{$\omega$(r)}	&	\multicolumn{1}{c}{v(r)}	\\
	&	\multicolumn{1}{c}{[km/s]}			&		&		&	\multicolumn{1}{c}{[\degr]} 	&	\multicolumn{1}{c}{[\degr]} 	&	\multicolumn{1}{c}{[pc]}	&	\multicolumn{1}{c}{[log yr]}	&		&	\multicolumn{1}{c}{[kpc]}	&	\multicolumn{1}{c}{[km/(s$\cdot$kpc)]} & \multicolumn{1}{c}{[km/s]}	\\
\hline
Berkeley 20	&	75.51	 $\pm$ 	4.85	&	9	&	a	&	203.483	&	-17.373	&	8710	&	9.763	&	a	&	16.0	 $\pm$ 	1.7 &	 10.6 $\pm$ 1.6	&	 169.5	 $\pm$ 	43.6 \\
Berkeley 25	&	134.30 $\pm$ 	1.62	&	4	&	e	&	226.612	&	-9.700	&	11400	&	9.699	&	e	&	17.7	 $\pm$ 	2.2	&	  9.6 $\pm$	0.3	&	 170.4 $\pm$ 26.4	\\
Berkeley 31	&	55.80	 $\pm$ 	1.13	&	2	&	o	&	206.254	&	5.120	&	8300	&	9.301	&	e	&	15.8	 $\pm$ 1.7	&	18.3 $\pm$ 	0.3	&	 290.4	 $\pm$ 	36.1	\\
Berkeley 32	&	105.00	 $\pm$ 	1.40	&	9	&	m	&	207.950	&	4.400	&	3162	&	9.720	&	n	&	10.9	 $\pm$ 	0.6	&	 5.9 $\pm$ 	0.4	&	 64.3	 $\pm$ 	~~7.8	\\
Berkeley 66	&	-50.65	 $\pm$ 	0.07	&	2	&	a	&	139.434	&	0.218	&	4570	&	9.580	&	a	&	11.9	 $\pm$ 	0.9	&	20.2 $\pm$ 	0.0	&	239.1	 $\pm$ 	17.6	\\
Berkeley 73	&	95.70	 $\pm$ 	0.57	&	2	&	e	&	215.278	&	-9.424	&	9800	&	9.176	&	e	&	16.8	 $\pm$ 	2.0	&	12.8 $\pm$ 	0.1	&	214.9	 $\pm$ 	27.0	\\
Berkeley 75	&	94.60	 $\pm$ 	0.35	&	1	&	e	&	234.307	&	-11.188	&	9100	&	9.602	&	e	&	15.1	 $\pm$  1.7	&	17.9 $\pm$ 	0.1	&	269.8	 $\pm$ 	31.1	\\
Cr 110	&	40.00	 $\pm$ 	1.00	&		&	d	&	209.650	&	-1.980	&	1950	&	9.230	&	d	&	9.7	 $\pm$ 	0.4	&	23.9 $\pm$ 	0.3	&	233.0	 $\pm$ 	11.4	\\
King 11	&	-35.00	 $\pm$ 	16.0	&		&	l	&	117.160	&	6.480	&	2198	&	9.615	&	n	&	9.2	 $\pm$ 	0.3	&	26.0 $\pm$ 	2.3	&	239.2	 $\pm$ 	28.4	\\
NGC 2243	&	61.00	 $\pm$ 	1.00	&		&	i	&	239.480	&	-18.010	&	3532	&	9.681	&	c, i	&	10.1	 $\pm$ 	0.5	&	23.4 $\pm$ 	0.2	&	236.8	 $\pm$ 	13.6	\\
NGC 2506	&	83.70	 $\pm$ 	1.40	&	4	&	f	&	230.560	&	9.940	&	3311	&	9.230	&	j, f	&	10.4	 $\pm$  0.6	&	 18.7 $\pm$ 	0.2	&	194.3	 $\pm$ 12.7	\\
NGC 6939	&	-18.98	 $\pm$ 	0.19	&	26	&	k	&	95.900	&	12.300	&	1820	&	9.114	&	b	&	8.4	 $\pm$ 	0.1	&	 29.1 $\pm$ 	0.0	&	244.0	 $\pm$ ~~3.6	\\
Pismis 2	&	49.20	 $\pm$ 	7.80	&	9	&	h	&	258.850	&	-3.340	&	3467	&	9.041	&	g	&	9.3	 $\pm$ 	0.4	&	 25.6	 $\pm$ 	1.0	&	 238.2	 $\pm$ 	19.3	\\
Saurer 1	&	104.60	 $\pm$ 	0.30	&	2	&	p	&	214.689	&	7.386	&	13200	&	9.699	&	e	&	20.2	 $\pm$ 2.7	&	10.2	 $\pm$ 	0.1	&	205.2	 $\pm$ 	28.6	\\
Tombaugh 2	&	120.51	 $\pm$ 	2.19	&	37	&	a	&	232.832	&	-6.880	&	7950	&	9.204	&	a	&	14.2	 $\pm$ 	1.5	&	 13.7	 $\pm$ 	0.3	&	 194.7	 $\pm$ 	25.2	\\
\hline
\end{tabular}
\caption{ Data of old open clusters and calculated velocities. If the error of radial velocity was not given we assumed 1 km/s.
  References:
  a	- \cite{Andreuzzi2011};
  b	- \cite{Andreuzzi2004};
  c	- \cite{Bonifazi};
  d	- \cite[and references therin]{Bragaglia};
  e	- \cite[and references therin]{Carraro2007};
  f	- \cite{Carretta};
  g	- \cite{Di Fabrizio};
  h	- \cite{Friel};
  i	- \cite{Gratton};
  j	- \cite{Marconi};
  k	- \cite{Milone};
  l	- \cite{Scott};
  m	- \cite{Sestito};
  n	- \cite{Tosi};
  o	- \cite{Yong};
  p	- \cite{Carraro2004}.
}
\label{tab_OC}
\end{table}
\end{landscape}

\section{Non-circular orbits}

   The source of discrepancy in rotation curve determinations may be the assumption of circular orbits.
  We have assumed elliptical orbits as the simplest model of non-circular orbits. 
  We checked if the assumption of stars on elliptical orbits is consistent with observed radial velocities.
  We have analyzed radial velocities of K and M giants in the Galactic anticenter from the CORAVEL spectrograph given by \cite{Coravel}.
Stars in binary systems have been removed from the analyzed sample.
Regardless of the size of square centered on the Galactic anticenter the standard deviations of radial velocities
can not be explained assuming circular orbits (Table \ref{StdDev}). We obtain standard deviation of radial
velocities similar to observed ones, assuming elliptical orbits with eccentricities uniformly distributed in the range 0--0.6 in the outer parts of our Galaxy. So, the assumption of non-circular orbits is consistent with observed radial velocities.

   We have made a Monte-Carlo simulation of stars on
elliptical orbits beyond the Sun -- Galactic center ($R_{\sun}$) distance.
The semi-major axis (a), eccentricity (e), true anomaly ($\nu$) and
argument of pericenter ($\omega$) were chosen randomly with uniform distribution.
All orbits were located in the Galactic plane. The mass inside the solar orbit was set to
$1.07\cdot10^{11}\ {\mathrm M}_\sun$, which corresponds to solar velocity $v_{\sun}=240$ km/s.

The rotation velocities were calculated using formulas, that were derived assuming circular orbits.
The rotation velocity was calculated from radial velocity ($v_\mathrm{r}$) of star using
\begin{equation} 
v(r)=\frac{r}{R_{\sun}} \left( \frac{v_\mathrm{r}}{\sin l} + v_{\sun} \right),   \label{vr}
\end{equation}
where $r$ is the distance between star and the Galactic center. 
We want to test the influence of formulas derived for circular orbits applied to stars on elliptical ones.

The semi-major axes in our simulation were distributed from 5 to 25 kpc to avoid truncation effects
at $R_{\sun}=8$ kpc. The stars with galactic longitudes less then 20\degr\ from 0\degr\ or 180\degr\
were not shown, because the denominator in eq. \ref{vr} is to small.
A small denominator in equation \ref{vr} 
leads to unphysically large rotation velocities, up to tens of thousands km per second. 

\begin{figure*}
    \centering
    \includegraphics[width=\textwidth]{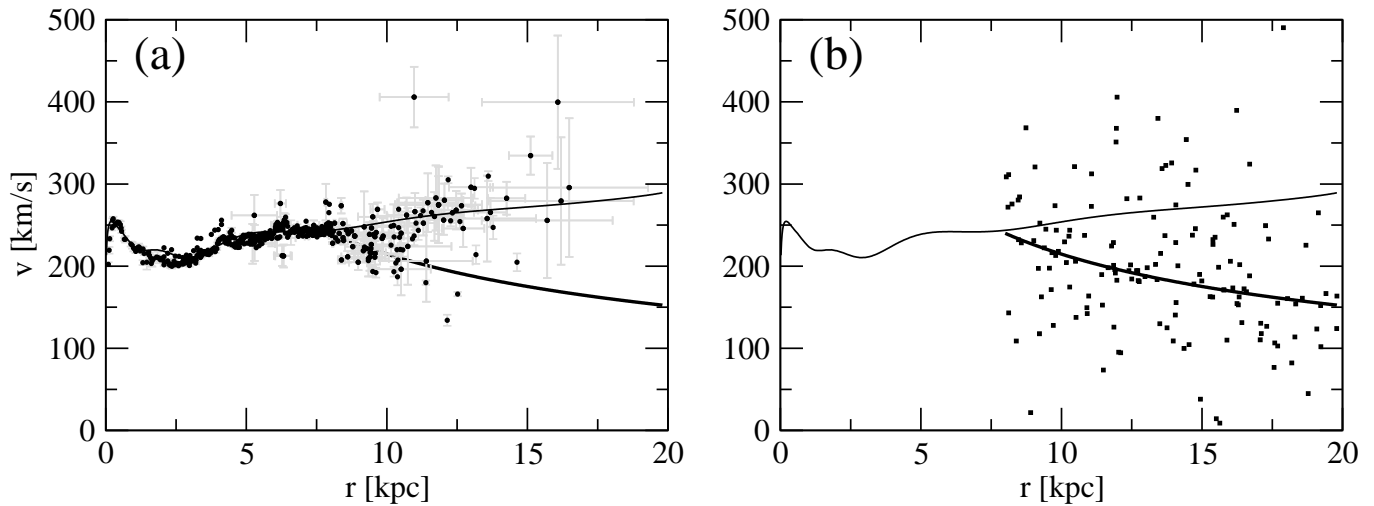}
    \caption{ Comparison of the Galaxy rotation curves: 
    {\bf (a)} Points from \protect\cite{Sofue2009} derived with tangent point method or from radial velocity. The \protect\cite{Sofue2009} data were recalculated using v$_\sun$=240~km/s. 
    {\bf (b)} Monte-Carlo simulation of stars on elliptical orbits with eccentricities e=0--0.5. The rotational velocities were calculated from radial velocities (eq. \ref{vr}).
    The lines show the model of flat rotation curve by \protect\cite{Sofue2009} and the Keplerian rotation curve (both for v$_\sun$=240~km/s)  .  }
    \label{rotVel}
 \end{figure*}

  The result of the Monte-Carlo simulation for 200 objects is shown on Fig. \ref{rotVel}(b).
Only objects with the distance 8--20 kpc from Galactic center are shown on the Monte--Carlo simulations plots.
The rotation velocities derived from radial velocities have large dispersion and look very similar to the observed rotation velocities from \cite{Sofue2009}.
The rotational velocities on Fig. \ref{rotVel}(b) are placed from below Keplerian rotation curve to above flat rotation curve.
This is similar to the rotational velocities in \cite{Sofue2009} compilation.

  The $\chi^2$ analysis was performed on the Monte-Carlo simulated points in the same way as with open clusters.
Although the simulated objects velocities were calculated assuming Keplerian rotation (on elliptical orbits e=0--0.5), the 
$\chi^2=1268$ for the agreement with Keplerian rotation curve is almost 8 times larger than $\chi^2_{1-0.005}$. 
The test of agreement with flat rotation curve $\chi^2=13201$ is almost 80 times larger than $\chi^2_{1-0.005}$.
So the rotation velocity derived from observed radial velocity in the case of highly eccentric orbits can
not be used to distinguish between flat and Keplerian rotation curves.

\section{Discussion}

The main argument for the flat rotation curve of our Galaxy, given by \citet{Sofue2009}, is the
observation of Sharpless 269 star forming region observed by VERA (VLBI Exploration of Radio Astrometry).
Both radial and transverse velocities \citep{VERA} lead to a rotation velocity of $\sim$200 km/s. 
Also the VLBI measurements of parallaxes and proper motions of star forming regions (SFR) published
by \citet{Reid} leads to flat rotation curve. 
It seems that SFR (maybe all objects in spiral arms) have different rotation velocity than interstellar clouds. 

The molecular clouds have the lowest velocity dispersion $\sigma_z=5$ km/s in the
direction perpendicular to the galactic plane, as compared to stars or \ion{H}{II} regions. 
Because of the large cross section they may be better thermalized than stars. 
Therefore the orbits of molecular clouds may have lower eccentricities than other objects. 
They match then very well the Keplerian rotation curve in our simulation, similar to velocities observed by \cite{Gazinur}.

\begin{table}
\centering
\caption{ Observed \citep{Coravel} and simulated (Monte-Carlo) standard deviations of radial velocity in Galactic anticenter. The square in which we analyze the radial velocities is centered on the Galactic anticenter. The minimum and maximum standard deviation of radial velocity is calculated from 10 Monte-Carlo simulations.}
\label{StdDev}
\begin{tabular}{rrrr}
\hline
                           & \multicolumn{1}{c}{observed} & \multicolumn{2}{c}{\underline{Monte-Carlo simulations}} \\
\multicolumn{1}{c}{square} & \multicolumn{1}{c}{std. dev.}   & \multicolumn{1}{c}{circular} & \multicolumn{1}{c}{elliptical} \\
\multicolumn{1}{c}{side}   & \multicolumn{1}{c}{of v$_r$} & \multicolumn{1}{c}{orbits}   & \multicolumn{1}{c}{orbits 0$\leq$e$\leq$0.6} \\
\multicolumn{1}{c}{[\degr]}& \multicolumn{1}{c}{[km/s]}   & \multicolumn{1}{c}{[km/s]}   & \multicolumn{1}{c}{[km/s]} \\ 
\hline
6	  & 38.0 &  3.1--6.2 &	31.4--51.6 \\
8	  & 43.7 &  4.1--7.1 &	28.7--56.7 \\
10	& 40.7 &  7.1--9.5 &	29.3--56.7 \\
12	& 39.6 &  8.0--11.8 &	34.0--50.8 \\
14	& 40.6 & 11.1--14.4 &	20.9--61.2 \\
16	& 39.0 & 12.1--15.2 & 36.8--48.5 \\
18	& 37.6 & 12.7--17.8 & 32.3--57.4 \\
20	& 37.6 & 15.4--18.6 & 35.6--53.9 \\
\hline
\end{tabular}
\end{table}

Eleven directions towards galactic anticenter were observed by \cite{Gazinur}.
The interstellar clouds are located in Galactic longitudes 184\degr$<$l$<$190\degr.
The radial velocities towards these clouds have a standard deviation of 1.8 km/s.
We cannot obtain such low dispersion in our simulations, even with circular orbits.
The standard deviation of radial velocity for circular orbits in the mentioned longitude range is 3.5--7.2 km/s.

\section{Conclusions}
  
  The rotation curve derived from observations of old open clusters seems
  to confirm the observations of Keplerian rotation curve for our Galaxy.
  The determination of flat rotation curve may be caused by applying the formula derived with the assumption of circular orbits to non-circular ones.
 The main results are:
\begin{itemize}
	\item The observations of flat or Keplerian rotation curve of our Galaxy can be explained
	assuming Keplerian rotation, elliptical orbits of stars and almost circular orbits of interstellar clouds.
	\item The Galactic rotation velocity derived from radial velocity in the case of elliptical orbits with high eccentricities can not be used	to distinguish between flat or Keplerian rotation curve. 
  \end{itemize} 

  The Keplerian rotation curve of the Galaxy will have a huge impact on the
  amount of dark matter in our Galaxy.
  


\begin{thebibliography}{}
    \bibitem[\protect\citeauthoryear{Andreuzzi {\it et~al.}}{2011}]{Andreuzzi2011}{Andreuzzi G., Bragaglia A., Tosi M., Marconi G.}, {2011}, {MNRAS}, {\bf 412}, {1265} 
    \bibitem[\protect\citeauthoryear{Andreuzzi {\it et~al.}}{2004}]{Andreuzzi2004}{Andreuzzi G., Bragaglia A., Tosi M., Marconi G.}, {2004}, {MNRAS}, {\bf 348}, {297}  
    \bibitem[\protect\citeauthoryear{Bhattacharjee {\it et~al.}}{2014}]{Bhattacharjee}{Bhattacharjee P., Chaudhury S., Kundu S.}, {2014}, {ApJ}, {{\bf 785}:63}
    \bibitem[\protect\citeauthoryear{Bonifazi {\it et~al.}}{1990}]{Bonifazi}{Bonifazi A., Tosi M., Fusi Pecci F., Romeo G.}, {1990}, {MNRAS}, {\bf 245}, {15}  
    \bibitem[\protect\citeauthoryear{Bragaglia {\it et~al.}}{2006}]{Bragaglia}{Bragaglia A., Tosi M.}, {2006}, {AJ}, {\bf 131}, {1544}  
    \bibitem[\protect\citeauthoryear{Carraro}{1994}]{Carraro1994}{Carraro G., Chiosi C.}, {1994}, {A\&A}, {\bf 288}, {751}
    \bibitem[\protect\citeauthoryear{Carraro {\it et~al.}}{2004}]{Carraro2004}{Carraro G., {\it et al.}}, {2004}, {AJ}, {\bf 128}, {1676}   
    \bibitem[\protect\citeauthoryear{Carraro {\it et~al.}}{2007}]{Carraro2007}{Carraro G., {\it et al.}}, {2007}, {A\&A}, {\bf 476}, {217}   
    \bibitem[\protect\citeauthoryear{Carretta {\it et~al.}}{2004}]{Carretta}{Carretta E., Bragaglia A., Gratton R., Tosi M.}, {2004}, {A\&A}, {\bf 422}, {951} 
    \bibitem[\protect\citeauthoryear{Demers \& Battinelli}{2007}]{Demers2007}{Demers S., Battinelli P.}, {2007}, {A\&A}, {\bf 473}, {143} 
    \bibitem[\protect\citeauthoryear{Di Fabrizio {\it et~al.}}{2001}]{Di Fabrizio}{Di Fabrizio L., Bragaglia A., Tosi M., Marconi G.}, {2001}, {MNRAS}, {328}, {795} 
    \bibitem[\protect\citeauthoryear{Famaey {\it et~al.}}{2005}]{Coravel}{ Famaey B., Jorissen A., Luri X., Mayor M. {\it et~al.}}, {2005}, {A\&A}, {\bf 430}, {165}  
    \bibitem[\protect\citeauthoryear{Fich {\it et~al.}}{1989}]{Fich1989}{Fich M., Blitz L., Stark A.}, {1989}, {ApJ}, {\bf 342}, {272}  
    \bibitem[\protect\citeauthoryear{Friel {\it et~al.}}{2002}]{Friel}{Friel E., {\it et al.}}, {2002}, {AJ}, {\bf 124}, {2693}  
    \bibitem[\protect\citeauthoryear{Galazutdinov {\it et~al.}}{2015}]{Gazinur}{Galazutdinov G., Strobel A. {\it et~al.}}, {2015}, {PASP}, {\bf 127}, {126}
    \bibitem[\protect\citeauthoryear{Gratton {\it et~al.}}{1994}]{Gratton}{Gratton R., Contarini G.}, {1994}, {A\&A}, {\bf 283}, {911} 
    \bibitem[\protect\citeauthoryear{Guetter}{1993}]{Guetter1993}{Guetter H.}, {1993}, {AJ}, {\bf 106}, {220} 
    \bibitem[\protect\citeauthoryear{Honma {\it et~al.}}{2007}]{VERA}{Honma M., Bushimata T., Choi Y.K. {\it et~al.}}, {2007}, {PASJ}, {\bf 59}, {889}
    \bibitem[\protect\citeauthoryear{Honma {\it et~al.}}{2012}]{Honma2012}{Honma M., Nagayama T., Ando K. {\it et~al.}}, {2012}, {PASJ}, {\bf 64}, {136}
    \bibitem[\protect\citeauthoryear{Honma {\it et~al.}}{2015}]{Honma2015}{Honma M. {\it et~al.}}, {2015}, {PASJ}, {\bf 67}, {70}
    \bibitem[\protect\citeauthoryear{Honma \& Sofue}{1997}]{Honma1997}{Honma M., Sofue Y.}, {1997}, {PASJ}, {\bf 49}, {539}
    \bibitem[\protect\citeauthoryear{Janes \& Phelps}{1994}]{Janes1994}{Janes K., Phelps R.}, {1994}, {AJ}, {\bf 108}, {1773}
    \bibitem[\protect\citeauthoryear{Marconi {\it et~al.}}{1997}]{Marconi}{Marconi G., Hamilton D., Tosi M., Bragaglia A.}, {1997}, {MNRAS}, {\bf 291}, {763}  
    \bibitem[\protect\citeauthoryear{Milone}{1994}]{Milone}{Milone A.}, {1994}, {PASP}, {\bf 106}, {1085}  
    \bibitem[\protect\citeauthoryear{Moni Bidin {\it et~al.}}{2012}]{Bidin}{Moni Bidin C., Carraro G., M\'endez R.A., Smith R.}, {2012}, {ApJ}, {\bf 751}, {30}
    \bibitem[\protect\citeauthoryear{Reid {\it et~al.}}{2014}]{Reid}{Reid M.J, Menten K.M. {\it et~al.}},{ApJ}, {\bf 783}, {130} 
    \bibitem[\protect\citeauthoryear{Scott {\it et~al.}}{1995}]{Scott}{Scott J., Friel E., Janes K.}, {1995}, {AJ}, {\bf 109}, {1706}  
    \bibitem[\protect\citeauthoryear{Sestito {\it et~al.}}{2006}]{Sestito}{Sestito P. {\it et al.}}, {2006}, {A\&A}, {\bf 458}, {121}   
    \bibitem[\protect\citeauthoryear{Sch\"onrich {\it et~al.}}{2010}]{Schonrich}{Sch\"onrich R., Binney J., Dehnen W.}, {2010}, {MNRAS}, {\bf \bf 403}, {1829}
    \bibitem[\protect\citeauthoryear{Sofue {\it et~al.}}{2009}]{Sofue2009}{Sofue Y., Honma M., Omodaka T.}, {2009}, {PASJ}, {\bf 61}, {227}
    \bibitem[\protect\citeauthoryear{Sofue}{2016}]{Sofue2016}{Sofue Y.}, {2016}, {arXiv:1608.08350}
    \bibitem[\protect\citeauthoryear{Tosi {\it et~al.}}{2007}]{Tosi}{Tosi M., Bragaglia A., Cignoni M.}, {2007}, {MNRAS}, {\bf 378}, {730}  
    \bibitem[\protect\citeauthoryear{Tosi {\it et~al.}}{1991}]{Tosi91}{Tosi M., Marconi G.L., Focardi P.}, {1991}, {AJ}, {\bf 102}, {951} 
    \bibitem[\protect\citeauthoryear{Yong {\it et~al.}}{2005}]{Yong}{Yong D., Carney B., Teixera de Almeida M.L.}, {2005}, {AJ}, {\bf 130}, {597}  
 

\end{thebibliography}
\end{document}